\pdfoutput=1
\documentclass[11pt]{article}

\usepackage[margin=1.4in]{geometry}
\usepackage{amsmath}
\usepackage{graphicx}
\usepackage{authblk}
\begin{document}

\title{\vspace{-0.6in} \bf \Large Walk modularity and community structure in networks}

\author[1]{David Mehrle}
\author[2]{Amy Strosser}
\author[3]{Anthony Harkin}
\affil[1]{Department of Mathematical Sciences, Carnegie Mellon University, Pittsburgh, PA 15213}
\affil[2]{Department of Mathematics and Computer Science, Mount St. Mary's University, Emmitsburg, MD 21727}
\affil[3]{School of Mathematical Sciences, Rochester Institute of Technology, Rochester, NY 14623}
\renewcommand\Authands{ and }
\renewcommand\Affilfont{\itshape\small}

\date{\vspace{-5ex}}

\renewcommand{\abstractname}{\vspace{-0.4in}}

\maketitle

\begin{abstract}
Modularity maximization has been one of the most widely used approaches 
in the last decade for discovering community structure in networks of 
practical interest in biology, computing, social science, statistical 
mechanics, and more. Modularity is a quality function that measures the 
difference between the number of edges found within clusters minus the 
number of edges one would statistically expect to find based on random 
chance. We present a natural generalization of modularity based on the 
difference between the actual and expected number of walks within clusters, 
which we call {\it walk-modularity}. Walk-modularity can be expressed in 
matrix form, and community detection can be performed by finding leading
eigenvectors of the walk-modularity matrix. We demonstrate community 
detection on both synthetic and real-world networks and find that 
walk-modularity maximization returns significantly improved results 
compared to traditional modularity maximization.
\end{abstract}

\section{Introduction}

The problem of detecting community structure within networks has received
considerable attention over the last decade, largely due to the rapidly 
increasing accessibility of large, real-world network data sets in many 
fields, including biology, social sciences, statistical mechanics, citation 
networks, and many more 
\cite{Girvan:2002, Good:2010, Lusseau:2003, Porter:2009, Reichardt:2006}. 
A community, or cluster, may intuitively be thought of as 
a collection of nodes more densely connected to each other than with 
nodes outside the cluster. In real-world networks derived from biological, 
social, and internet data for example, one often does not know in advance 
how many separate communities are within the network, if any at all, nor 
how many nodes comprise each of the communities.

Several popular and effective algorithms for community detection in networks 
rely on the concept of modularity
\cite{Agarwal:2008,  Barber:2007, Blondel:2008, Newman:2006,Newman:2006a, Newman:2004}.
The modularity function was initially designed as a quantitative measure 
of the quality of a given partition of a network \cite{Girvan:2002}. 
The sensible observation motivating the modularity measure is that separate 
communities should have significantly fewer edges running between 
them than one would expect to find based on random chance. Equivalently, 
the number of edges connecting nodes within a community should be far 
greater than statistically expected. The modularity function is therefore 
defined as the difference between the edge density within groups of nodes 
minus the expected edge density of an equivalent network with the edges 
placed at random.  Large, positive values of modularity indicate the presence 
of a good partition of a network into communities, and many community detection 
algorithms are based on searching for partitions which maximize the modularity 
function. 

In this work, we propose a natural generalization of modularity, 
that we call walk-modularity.  A walk on a network begins at a starting 
vertex and continues going along edges, which may be repeated, until it 
reaches a specified last vertex. The number of edges traversed is called 
the length of the walk. The modularity generalization we propose is that 
the number of walks of a specified length connecting nodes within a community 
should be far greater than expected by random chance. Large, positive values 
of the walk-modularity function will indicate groups of nodes that are very 
tightly knit together. For walk length equal to one, the walk-modularity 
function reduces to the standard modularity measure, which just considers
edges, and we refer to it as edge-modularity. 

In the seminal paper of Newman \cite{Newman:2006}, the modularity function 
was cast into a matrix formulation, yielding the so-called modularity matrix 
of a network. The modularity function can be maximized by computing leading 
eigenvectors of the modularity matrix, and high quality network partitions 
are obtained. For any specified walk length, we can define the walk-modularity 
matrix and can use it to find a good partition of a network into communities.
The method of optimal modularity has several desirable features which have 
led to its widespread popularity. In particular, neither the number or sizes 
of the communities has to be specified in advance.  These desirable 
features of the method of optimal modularity are retained when using the
walk-modularity generalization.

\section{Walk-Modularity}

Two nodes in a network are connected by a length $\ell$ walk if you can
start at one node and traverse $\ell$ consecutive edges to reach the other 
node.  The generalized notion of modularity we propose is based on the 
idea that a community will have a statistically unexpected number of 
walks between nodes. We therefore express the walk-modularity as
\begin{equation*}
\begin{split}
Q_{\ell} & = \text{(number of walks of length $\ell$ within communities)} \\
        & \qquad \text{- (expected number of such walks).}
\end{split}
\end{equation*}
Partitions of a network with large positive values of $Q_{\ell}$ will indicate 
groups of nodes that are very tightly knit together. 
Suppose a network contains $n$ nodes and has an adjacency matrix denoted by $A$.
It is a well known fact that powers of the adjacency matrix can be used
to count the numbers of walks between pairs of nodes in a graph.  The $i,j$
entry of the matrix $A^{\ell}$ is the number of walks of length $\ell$ between 
vertices $i$ and $j$.

In order to calculate the expected number of walks, we need to consider 
an equivalent randomized network with which to compare the real network.
The expected number of walks between vertices will depend on the choice of 
the so-called null model to which we compare our network. We will use the
same null model described in the seminal paper of Newman \cite{Newman:2006}.
In that work, an equivalent null model was considered where edges are placed
entirely at random, subject to the constraint that the expected degree of
each vertex in the null model is the same as the actual degree of the
corresponding vertex in the real network. Let $k_{i}$ be the degree of vertex
$i$ and let $m$ be the total number of edges in the network.  The expected 
number of edges between vertices $i$ and $j$ is then given by
\begin{equation*}
\label{expected edges}
P_{ij} = \frac{k_{i}\,k_{j}}{2m} .
\end{equation*}
Powers of the expected adjacency matrix, $P$, will count the expected
number of walks in the null model.  The $i,j$ entry of the matrix $P^{\ell}$ 
is the expected number of walks of length $\ell$ between vertices $i$ and $j$.

Define $g_{i}$ to be the community to which vertex $i$ belongs.
The walk-modularity, $Q_\ell$, for walks of length $\ell$ can be written
\begin{equation*}
\label{walk modularity}
Q_\ell = \frac{1}{2 m_\ell}\sum_{i,j} \big[ (A^\ell)_{ij} - (P^\ell)_{ij} \big] 
\delta(g_{i}, g_{j})\ ,
\end{equation*}
where $\delta(r,s) = 1$ if $r = s$ or zero otherwise. 
Walk-modularity only compares the actual number of walks versus the expected 
number of walks for vertices within the same community. We choose to normalize
walk-modularity by  
\begin{equation*}
m_\ell = \frac{1}{2} \sum_{i,j} (A^\ell)_{ij}\ .
\end{equation*}
Note that for $\ell = 1$, the walk-modularity $Q_{1}$ reduces to the standard
definition of modularity given in \cite{Girvan:2002}, which we refer to as 
edge-modularity. The walk length, $\ell$, will turn out to be a natural parameter 
in the clustering algorithm that provides the ability to control how tightly knit 
a community of nodes should be.

\section{Spectral Optimization of Walk-Modularity}
\label{Spectral Maximization}

To find community structure in a network, we wish to maximize the 
walk-modularity over all possible partitions of the network. A large 
number of community detection techniques have been based on maximizing 
edge-modularity, such as in \cite{Agarwal:2008,Blondel:2008,Brandes:2007,Newman:2004}, 
although \cite{Brandes:2007} proved that the decision problem of finding a
clustering with modularity exceeding a given value is {\bf NP}-complete.
Clustering algorithms attempting to maximize modularity focus on
heuristics and approximations to approach the optimum value.
There are many heuristic algorithms for maximizing edge-modularity, 
using techniques such as linear programming, vector programming, 
greedy agglomeration approaches, extremal optimization, and spectral optimization
\cite{Agarwal:2008, Arenas:2005, Blondel:2008, Brandes:2007, Newman:2006,Newman:2006a}.
In what follows, we employ the spectral optimization procedure from 
\cite{Newman:2006} adapted to work for walk-modularity, which proved 
to be the simplest optimization technique to implement while giving 
very high quality results.

Closely following the discussion in \cite{Newman:2006,Newman:2006a},
we first consider the problem of partitioning the network into just
two groups of nodes. Define a length $n$ index vector 
${\bf s}$ such that
\begin{equation*}
\label{association vector}
s_{i} =
\begin{cases}
  +1 & \text{if vertex $i$ belongs to group 1,} \\
  -1 & \text{if vertex $i$ belongs to group 2.}
\end{cases}
\end{equation*}
Noting that the quantity $\delta(g_{i},g_{j})=\frac{1}{2} (s_i s_j + 1)$ 
is $1$ if nodes $i$ and $j$ are in the same group, and 0 otherwise, 
we can write the walk-modularity as
\begin{align*}
\label{rewriting walk modularity 1}
Q_\ell & = \frac{1}{4m_\ell} \sum_{i,j}  \big[ (A^\ell)_{ij} - (P^\ell)_{ij} \big] (s_i s_j + 1) \\
      & = \frac{1}{4m_\ell} \sum_{i,j}  \big[ (A^\ell)_{ij} - (P^\ell)_{ij} \big] s_i s_j 
           +  \frac{1}{4m_\ell} \sum_{i,j}  \big[ (A^\ell)_{ij} - (P^\ell)_{ij} \big]
\end{align*}
If we define the walk-modularity matrix as 
\begin{equation*}
B=A^{\ell}-P^{\ell}
\end{equation*}
then
\begin{equation}
\label{Ql}
Q_\ell  = \frac{1}{4m_\ell} {\bf s}^{T}B{\bf s} + \frac{1}{4m_\ell} \sum_{i,j} B_{ij}
\end{equation}

Since large, positive values of walk-modularity indicate the presence of a good division
of the network, the goal is to obtain the index vector ${\bf s}$ which maximizes the 
value of $Q_\ell$. This amounts to optimizing ${\bf s}^T B {\bf s}$
since the second term in (\ref{Ql}) does not depend on the choice of ${\bf s}$.
Let $\lambda_1 \geq \lambda_2 \geq \ldots \geq \lambda_n$ be the eigenvalues of
$B$, with associated orthonormal eigenvectors ${\bf u}_1, {\bf u}_2, \ldots, {\bf u}_n$. 
Expressing ${\bf s}$ as a linear combination of the eigenvectors 
${\bf s} = \sum_{i=1}^n a_i {\bf u}_i$ with $a_i = {\bf u}_i^T {\bf s}$,
we obtain
\begin{equation*}
{\bf s}^T B {\bf s} \
= \ \left(\sum_{i=1}^n a_i {\bf u}_i^T\right) B
  \left( \sum_{i=1}^n a_i {\bf u}_i \right) \
= \ \sum_{i=1}^n a_i^2 \lambda_i .
\end{equation*}
Hence, the task of choosing ${\bf s}$ to maximize $Q_\ell$ is
equivalent to choosing positive values $a_i^2$ to place as much weight
as possible on the term in the sum with the largest positive eigenvalue, 
namely $\lambda_1$. Since $a_1 = {\bf u}_1^T {\bf s}$, we choose ${\bf s}$ to 
be as close to parallel with ${\bf u}_1$ as possible. Thus, as in 
\cite{Newman:2006}, the choice of ${\bf s}$ is
\begin{equation*}
\label{choosing s}
s_i = \begin{cases}
  +1 & u^{(1)}_i \geq 0 \\
  -1 & u^{(1)}_i < 0
\end{cases} \hspace{1em},
\end{equation*}
where $u^{(1)}_i$ is the $i$-th entry of ${\bf u}_1$. Therefore, nodes are 
placed into either group 1 or group 2 depending on the sign of their 
corresponding entry in the leading eigenvector of the walk-modularity matrix. 
If the overall value of walk-modularity is $Q_{\ell} \leq 0$ then the
network should not be partitioned in two.

In order to divide a network into more than just two communities, we will
employ a recursive approach where we keep dividing groups in two until we 
find indivisible communities. In order to decide whether a particular group 
should be further divided, we must examine the change in walk-modularity that
would result. Proceeding again as in \cite{Newman:2006}, we consider a group
$g$ with $n_{g}$ nodes, and calculate the change in walk-modularity that would 
result from further division of the group into two pieces,
\begin{eqnarray*}
\label{change in modularity}
\Delta Q_\ell &=& \frac{1}{2m_{l}} \left[
                     \frac{1}{2}\sum_{i,j \in g} B_{ij} (s_{i}s_{j}+1)
                     - \sum_{i,j \in g} B_{ij} \right] \notag \\
              &=& \frac{1}{4m_{l}} \left[
                  \sum_{i,j \in g} B_{ij} s_{i}s_{j}
                  - \sum_{i,j \in g} B_{ij} \right] \notag \\
	      &=&  \frac{1}{4m_{l}} \sum_{i,j \in g} \left[
                     B_{ij} - \delta_{ij} \sum_{k \in g} B_{ik} \right]
                     s_{i}s_{j} \notag \\
              &=&  \frac{1}{4m_{l}} {\bf s}^{T}B^{(g)}{\bf s} \label{deltaQL} 
\end{eqnarray*}
In a recursive approach for dividing a network into multiple communities,
maximizing the contribution to walk-modularity from subdivision of communities 
can be approached using the same leading eigenvector method as before, but using 
the matrix $B^{(g)}$ at each step.  Furthermore, a recursive community detection 
algorithm should refuse to make any subdivisions for which the change in 
walk-modularity is negative, which can be determined by explicitly calculating 
the value of $\Delta Q_\ell$ at each step. Communities for which 
$\Delta Q_{\ell} \leq 0$ are called indivisible. In this manner, the spectral 
approach can be applied to divide a network into multiple indivisible communities 
without the need to specify in advance the number or sizes of the communities.

\section{Examples}
\label{Examples}

To demonstrate that walk-modularity is a very effective generalization of
edge-modularity for finding communities, we applied the spectral optimization
algorithm described in the previous section to both computer generated test
cases and real-world networks. It is very important to emphasize that, because 
our focus is conducting a consistent comparison between edge-modularity and 
walk-modularity, only the leading eigenvector algorithm is used for the 
tests in sections \ref{Synthetic Networks} and \ref{Real-World Networks}. 
No other enhancements for modularity maximization were included in this work.
The only parameter which varies from test to test is the walk length $\ell$.
The real-world test networks examined in section 4.2 are culled from the data 
in \cite{Lusseau:2003, Zachary:1977}. These two networks are commonly used as 
examples of real-world networks in community-detection literature
\cite{Agarwal:2008, Arenas:2008, Arenas:2005, Blondel:2008, Brandes:2007, 
Girvan:2002, Newman:2006, Newman:2006a, Newman:2004}.
The synthetic test networks in section 4.1 are randomly generated using either 
a variation on the standard Erd\H{o}s-R\'{e}nyi model of random networks or 
generated following the benchmark test described in \cite{Lancichinetti:2008}, 
which has previously been used to benchmark community-detection algorithms in
\cite{Agarwal:2008, Berry:2009, Fortuano:2009, Newman:2011}, among others.

\subsection{Synthetic Networks}
\label{Synthetic Networks}

\begin{figure}[!b]
  \centering
  \includegraphics[scale=0.35,angle=0,origin=c]{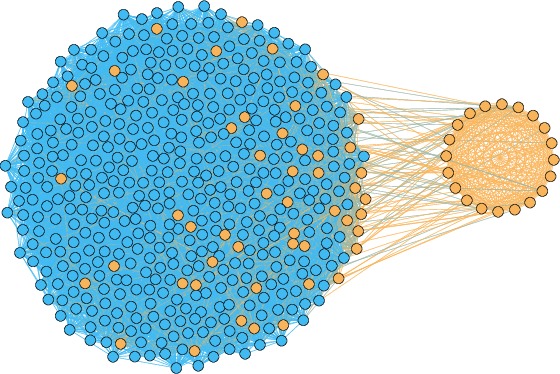}
  \hspace*{0.0in}
  \includegraphics[scale=0.35,angle=0,origin=c]{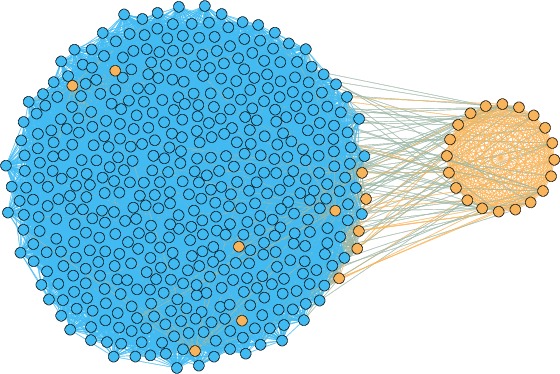}
  \\[-0.0in]
  \hspace{0.0in} $\ell = 1$ \hspace{2in} $\ell = 2$ \\[0.2in]
  \includegraphics[scale=0.35,angle=0,origin=c]{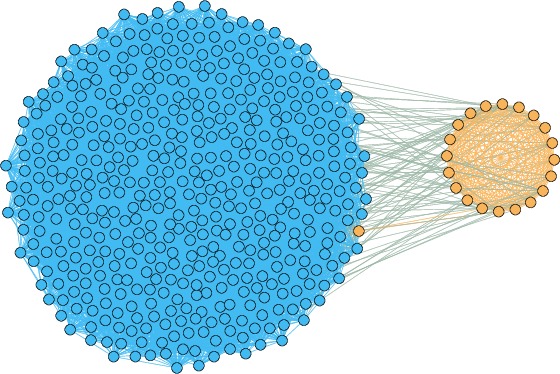}
  \hspace*{0.0in}
  \includegraphics[scale=0.35,angle=0,origin=c]{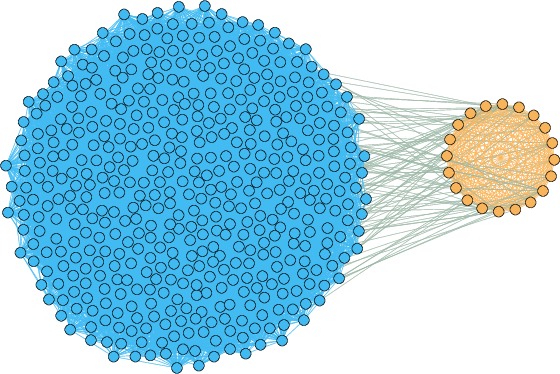}
  \\
  \hspace{0.0in} $\ell = 3$ \hspace{2in} $\ell = 4$ \\[0.2in]
  \caption{Random network $(n=500)$ with an embedded $K_{20}$. 
    The two communities found by edge-modularity, $\ell = 1$, 
    have $50$ misplaced nodes (top left).
    The two communities found by walk-modularity with $\ell = 2$ 
    have $11$ misplaced nodes (top right). 
    When $\ell = 3$ there is only $1$ misplaced node (bottom left),
    and when $\ell = 4$ every single node is placed correctly (bottom right).
    \label{embedk20mod}}
\end{figure}

Walk-modularity significantly outperforms edge-modularity on the
synthetic networks we examine in this section.
The first synthetic test shown is a modified Erd\H{o}s-R\'{e}nyi random
network with $n=500$ nodes with a $K_{20}$ complete subgraph connected to
it (Fig.~\ref{embedk20mod}).
The probability of an edge between two nodes in the random network seen in 
figure~\ref{embedk20mod} is $0.10$. The probability of an edge connecting 
a node in the $K_{20}$ subgraph to a node in the random network is $0.5$. 
The average degree of nodes in the complete subgraph are roughly the same 
as the average degree of nodes in the random network. To consider a node 
as either misplaced or placed correctly, the two communities were defined
to be the embedded $K_{20}$ and the Erd\H{o}s-R\'{e}nyi random network.
If the community detection algorithm placed a node from the complete subgraph 
in the same community as the rest of the network, or vice versa, the node 
is counted as misplaced. In other words, we are trying to compare how well 
the algorithms can find an embedded complete subgraph. In this test, we find
that walk-modularity significantly outperforms edge-modularity in that simply 
increasing the length of walks considered from $\ell = 1$ to $\ell = 3$ results 
in a decrease from $50$ misplaced nodes to merely $1$ misplaced node. When 
$\ell = 4$, walk-modularity correctly places every single node into the
two communities.

\begin{figure}
  \centering
  \includegraphics[scale=.4]{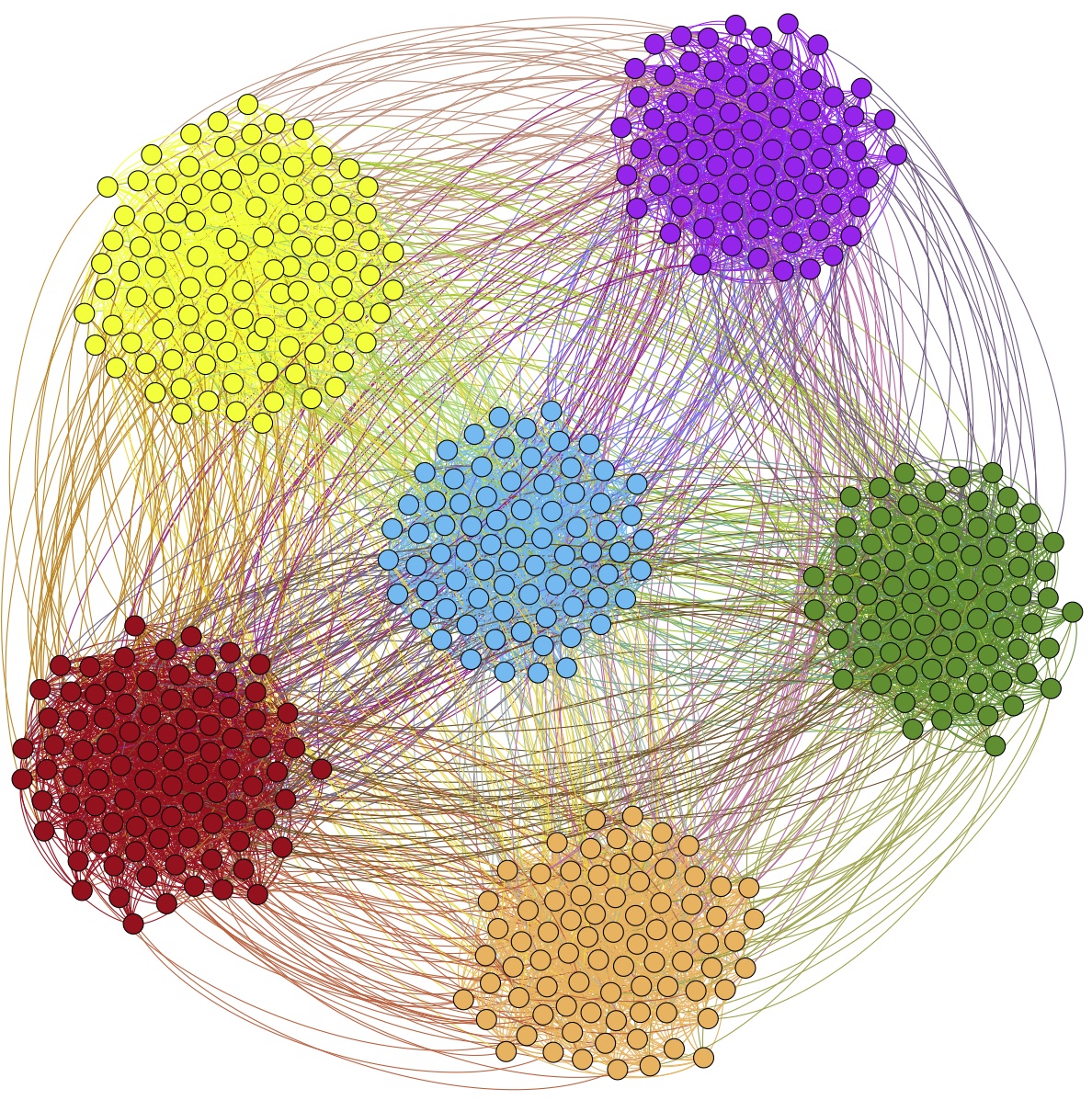}
  \caption{Communities as defined by the benchmark test with parameters
    $n = 500, \mu = 0.15, \bar k = 25$. \label{benchmark}}
\end{figure}

\begin{figure}
  \centering
  \includegraphics[scale=.4]{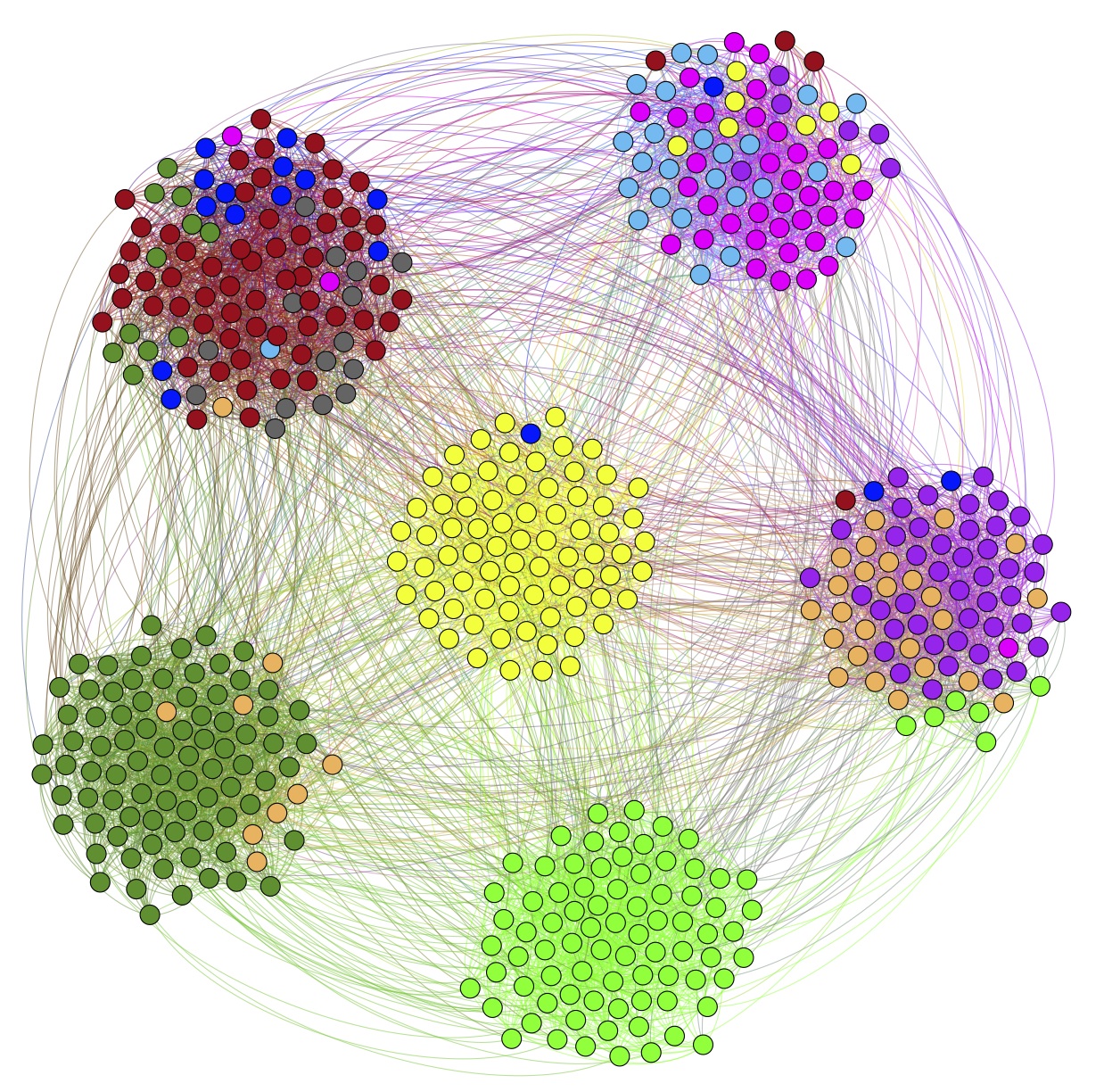}
  \caption{Communities as detected by traditional edge-modularity, $\ell = 1$. 
    \label{benchmarkmod}}
\end{figure}

\begin{figure}[h!]
	\centering
  \includegraphics[scale=.4]{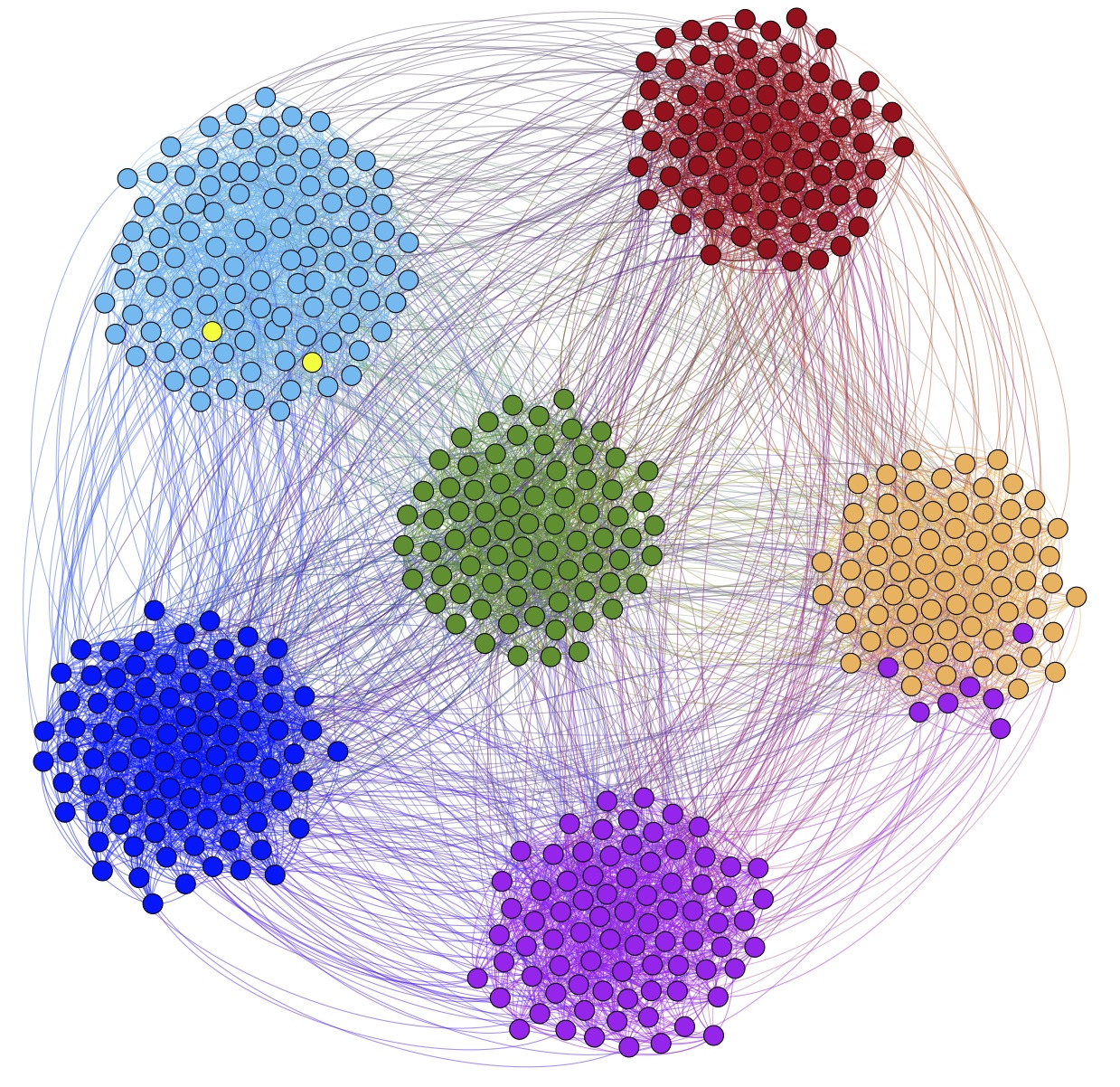}
  \caption{Communities as detected by walk-modularity with $\ell = 8$. 
    \label{benchmarkwalk8}}
\end{figure}

We next demonstrate the benchmark tests from \cite{Lancichinetti:2008}.
This synthetic test simulates a real-world network by placing each node in 
a well-defined community following a user-specified average degree, $\bar k$, 
and then randomly rewires nodes between communities according to a mixing 
parameter $\mu$. The result is a network with several communities with an 
approximate proportion of $1 - \mu$ edges among each community and $\mu$ 
edges between any one community and the others. 
One such network is pictured in figure \ref{benchmark}, for
parameters $n = 500, \mu = 0.15$, and $\bar k = 25$. Again, walk-modularity
significantly outperforms edge-modularity in this test case, although the number 
of nodes misplaced is somewhat difficult to quantify due to the variable number 
of communities that may be found. Figure \ref{benchmark} shows the communities 
as they were originally defined by the benchmark test. Figure~\ref{benchmarkmod} 
shows the communities found by edge-modularity, and figure~\ref{benchmarkwalk8} 
shows the result of the walk-modularity maximization with $\ell = 8$.

\subsection{Real-World Networks}
\label{Real-World Networks}

\begin{figure}
  \centering
  \includegraphics[scale=.55]{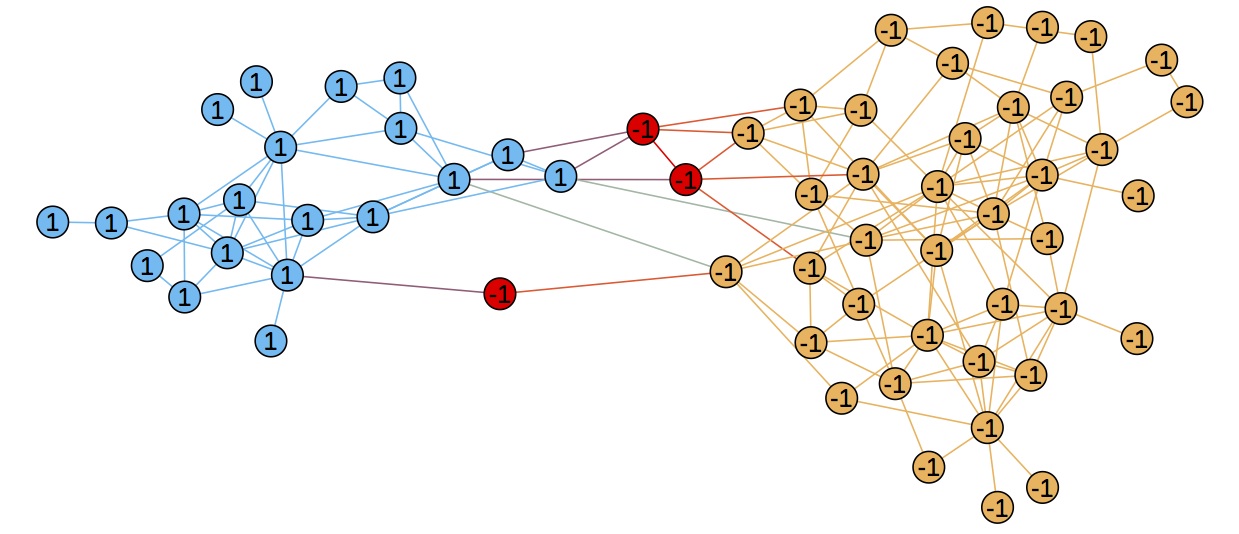}
  \vspace*{0.15in}
  \includegraphics[scale=.55]{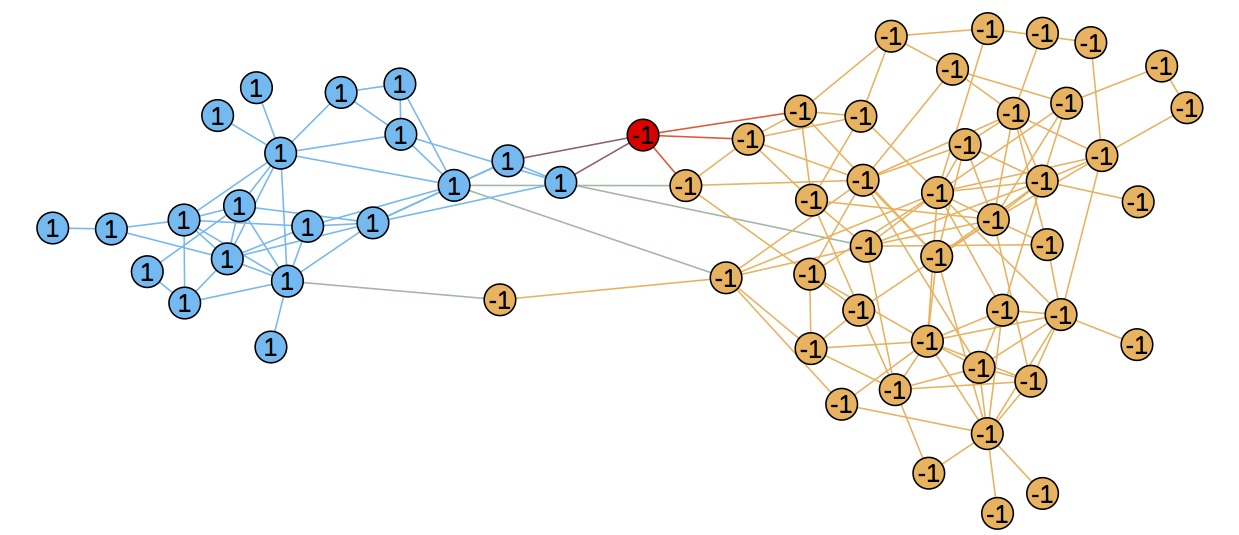}
  \caption{The dolphins network of \protect\cite{Lusseau:2003}. 
    Nodes are labeled with $\pm 1$ according to the observed separation 
    of the dolphin network by biologists. 
    (Top) Nodes are colored according to the partitioning found by 
    edge-modularity maximization. Misplaced nodes are colored red. 
    (Bottom) Nodes are colored according to the partitioning found by
    walk-modularity maximization with $\ell = 10$. There is only a 
    single misplaced node.
    \label{dolphinsmod}}
\end{figure}

The dolphins network from \cite{Lusseau:2003} is a social network of 62 dolphins
from Doubtful Sound, New Zealand, with edges representing social relations
between individuals, as established by observation over the course of a decade.
During the decade of observation in this study, the network of dolphins split 
into two separate communities. In a partition of the dolphins network, a node 
is considered misplaced if the algorithm places it into a different community 
than that observed by the biologists in \cite{Lusseau:2003}. Partitioning this 
network using edge-modularity finds $3$ nodes misplaced, as in figure 
\ref{dolphinsmod} (top).  After increasing the length of walks considered to 
$\ell = 8$, the diameter of this network, the number of nodes misplaced drops 
to merely $2$, and a value of $\ell = 10$ counts only a single misplaced node, 
figure \ref{dolphinsmod} (bottom).

\begin{figure}
  \centering
  \includegraphics[scale=.5]{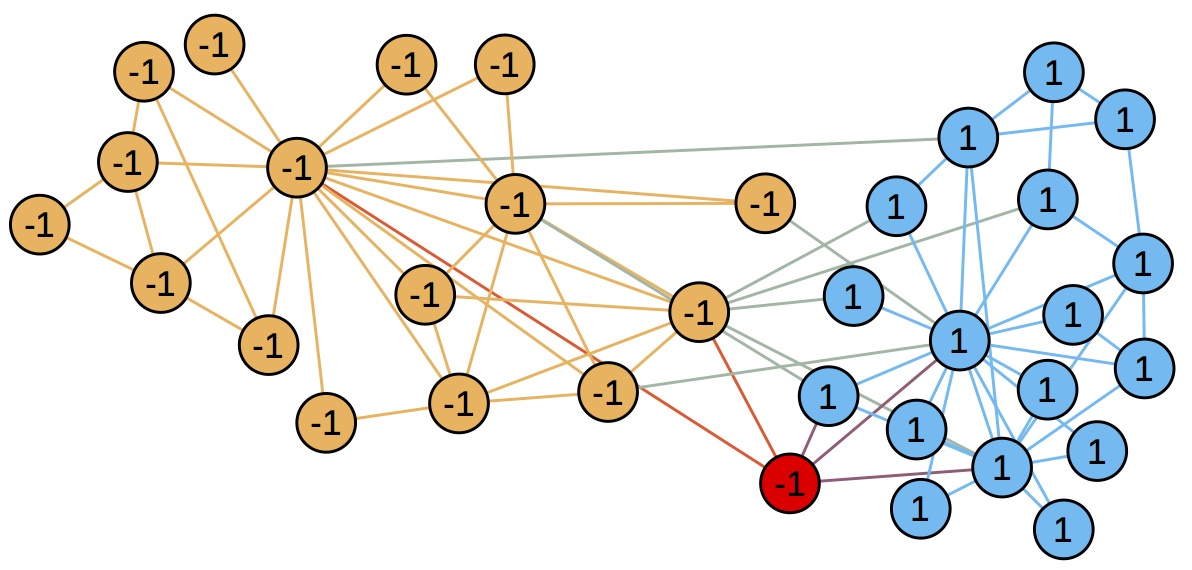}
  \caption{The karate club network of \protect\cite{Zachary:1977}. The nodes are
    colored relative to the partition found by the community detection algorithm,
    and the results are the same for both walk-modularity $\ell = 5$ and edge-
    modularity. The nodes are labelled $\pm 1$ based on the observed partition,
    and colored red to indicate a node that was incorrectly placed relative to
    the observed. \label{karateclub} }
\end{figure}

The karate club network from \cite{Zachary:1977} is another social network, this
time consisting of 34 members of a karate club, which once again split into two
communities during the course of observation following a disagreement between
the club's instructor and administrator. The nodes of this network represent
individuals, and the edges friendships as determined by Zachary. As before, 
a node is considered misplaced if the algorithm places it in a community which 
differs from the observed partition. In this case, there is no effect found from
increasing $\ell$, although the results are no worse than with edge-modularity.
The partition found from both walk-modularity with $\ell = 5$, the diameter of
this network, and edge-modularity are the same, as shown in figure~\ref{karateclub}.

For a given network, the user-controlled parameter $\ell$ can significantly impact 
the quality of the partition found by maximizing walk-modularity. In practice, we 
have found that choosing a value for $\ell$ somewhat near to the diameter of the 
network gives excellent results, as demonstrated in Table 1.  
Although it is typically computationally expensive to find the exact diameter of 
a network, there are several fast approximation algorithms
\cite{Crescenzi:2012, Roditty:2013}.
Moreover, we emphasize that even by using small values of $\ell > 1$, 
we obtained significant improvements in the observed quality of partitions.
We summarize the test case results of this section in Table 1.

\begin{table}
    \centering
    \begin{tabular}{c|c|c|c}
      Network & \ Diameter\, &\hspace*{1em} $\ell$ \hspace*{1em} &\ Nodes Misplaced \\ 
      \cline{1-4} 
      \rule{0in}{0.2in}
      Embedded $K_{20}$ & 3 & 1 & 50 \\
      & & 2 & 11 \\
      & & 3 & 1 \\ 
      & & 4 & 0 \\ 
      \cline{1-4}
      \rule{0in}{0.2in}
      Benchmark Test \cite{Lancichinetti:2008} & 4 & 1-3 & $>100$\\ 
      $n = 500$ & & 4 & 72 \\ 
      $\mu = 0.15$ & & 5 & 92 \\
      $\bar k = 25$ & & 6 & 40 \\ 
      & & 7 & 11 \\
      & & 8 & 9 \\ 
      \cline{1-4} 
      \rule{0in}{0.2in}
      Dolphins \cite{Lusseau:2003} & 8 & 1-7 & 3 \\
      & & 8, 9 & 2 \\
      & & 10 & 1 \\ 
      \cline{1-4}
      \rule{0in}{0.2in}
      Karate Club \cite{Zachary:1977} & 5 & 1-5 & 1
          \end{tabular}
  \caption{Summary of the number of nodes placed in incorrect communities
           as the walk length parameter, $\ell$, varies. In each case, as
           $\ell$ increases the community detection becomes more accurate.}
  \label{nodes misplaced}
\end{table}

\section{Conclusion}

We introduced a new measure of the quality of a partition of a network into 
communities, which we call walk-modularity. Walk-modularity is a natural
generalization of modularity because it considers the difference between 
the actual and expected number of walks of length $\ell$ within communities.
Mathematically, it is a very simple and elegant generalization in that
it only involves taking the $\ell$-th powers of both the adjacency matrix
and the expected adjacency matrix. As with traditional edge-modularity, 
walk-modularity can be maximized by finding the leading eigenvector of a 
matrix, called the walk-modularity matrix.  Although we have only explored,
in this paper, a single technique to maximize walk-modularity, it is a 
quantity that is compatible with maximization algorithms other than spectral 
maximization. 

Even with small values of $\ell > 1$, we demonstrated with test cases that maximizing 
walk-modularity produces partitions with many fewer misplaced nodes than traditional 
edge-modularity. For a random network with an embedded complete graph, $K_{20}$, 
walk-modularity is capable of perfectly identifying the complete subgraph and 
separating it from the random network. Walk-modularity is also more successful 
in identifying the six communities in a randomly generated benchmark test 
where edge-modularity did not perform well.  With two standard real-world 
community detection test cases, namely the dolphin network and the karate 
club network, walk-modularity performs in a manner comparable to the other 
most common community detection algorithms, and perhaps a bit better 
than edge-modularity in our comparison on the dolphin network.

\section{Acknowledgments}

This material is based upon work supported by the National Science Foundation 
under Grant No. DMS-1062128.

\bibliography{references}

\end{document}